\documentclass[letter]{aa} 
\usepackage{graphicx}

\newcommand{\rstar}{$R_{\star}$}
\usepackage{txfonts}
\begin{document}
   \title{Water in IRC+10216: a genuine formation process by shock-induced chemistry in the inner wind.}


   \author{I. Cherchneff
          }

   \institute{Departement Physik, Universit{\"a}t Basel,
              Klingelbergstrasse 82, CH-4056 Basel, Switzerland\\
              \email{isabelle.cherchneff@unibas.ch}
             }

   \date{Received October 31, 2010; accepted December 22, 2010}

 
  \abstract
   {The presence of water in the wind of the extreme carbon star IRC+10216 has been confirmed by the Herschel telescope. The regions where the high-J H$_2$O lines have been detected are close to the star at radii r $\leq$ 15 \rstar.}
   {We investigate the formation of water and related molecules in the periodically-shocked inner layers of IRC+10216 where dust also forms and accelerates the wind. }
   {We describe the molecular formation by a chemical kinetic network involving carbon-and oxygen-based molecules. We then apply this network to the physical conditions pertaining to the dust-formation zone which experiences the passage of pulsation-driven shocks between 1 and 5 \rstar. We solve for a system of stiff, coupled, ordinary, and differential equations.}
   {Non-equilibrium chemistry prevails in the dust-formation zone. H$_2$O forms quickly above the photosphere from the synthesis of hydroxyl OH induced by the thermal fragmentation of CO in the hot post-shock gas. The derived abundance with respect to H$_2$ at 5 \rstar\ is 1.4$\times 10^{-7}$, which excellently agrees the values derived from Herschel observations. The non-equilibrium formation process of water will be active whatever the stellar C/O ratio, and H$_2$O should then be present in the wind acceleration zone of all stars on the Asymptotic Giant Branch. }
   {}


   \maketitle
%

\section{Introduction}
The extreme carbon star IRC+10216 is one of the best studied evolved low-mass stars on the Asymptotic Giant Branch (AGB) owing to its proximity (d $\sim$ 180 pc) and the rich chemistry of its wind. Indeed, more than 60 molecular species have been detected at millimetre (mm), submillimetre (submm)  and infrared (IR) wavelengths, probing the entire gas conditions pertaining to the outflow (Ziurys \cite{zu06}). Because the star has already experienced third dredge-up, it is 'carbon-rich', i.e., its photosphere is characterised by a C/O ratio greater than 1. Many carbon-bearing species have been identified in the inner envelope extending from 1 to $\sim$ 10 \rstar, including CO, HCN, C$_2$H$_2$, CS, and SiC$_2$ at mid-IR and submm wavelengths (Keady \& Rigdway \cite{kea93}, Sch{\"o}ier et al. \cite{sho07}, Fonfr{\`i}a et al. 2008, Patel et al. \cite{pa09}, Cernicharo et al. \cite{cerni10}). Oxygen-bearing species other than CO have also long been detected in IRC+10216. Submm observations of silicon monoxide, SiO, confirmed a formation locus close to the star (Sch{\"o}ier et al. 2006, Decin et al. \cite{dec10}). The water molecule, H$_2$O, was first detected with the submm satellite SWAS by Melnick et al. (2001), and its presence was confirmed by submm observations with the ODIN satellite (Hasegawa et al. \cite{hag06}) and supported by the detection of hydroxyl, OH, by Ford et al. (2004). Recently, water was detected by the SPIRE, PACS, and HIFI spectrometers onboard the {\it Herschel} submm telescope in the carbon star VY Cyg (Neufeld et al. \cite{neu10}), the S star $\chi$ Cyg (characterised by a C/O ratio $\sim$ 1) (Justtanont et al. \cite{jus10}), and IRC+10216 (Decin et al. \cite{dec110}, hereafter DAB10). These various detections of high-excitation rotational lines probe relatively high gas temperatures, thus implying the presence of H$_2$O fairly close to the star. Derived abundances with respect to H$_2$ span values from 10$^{-8}$ to 10$^{-7}$ for IRC+10216, $\sim$ 10$^{-6}$ for VY Cyg, and 10$^{-5}$ for $\chi$ Cyg. The derived abundances seem to logically decrease as the carbon content of the star increases, because CO traps most of the oxygen available for water formation.
\begin{table*}
\caption{Gas parameters in the inner wind of IRC+10216. The shocks form at 1.2 \rstar. For each radius, the temperature and number density are given for the pre-shock gas, the gas after the shock front (in the collision-induced H$_2$ dissociation zone), and the gas at the beginning and the end of the adiabatic cooling zone ($\equiv$ ballistic trajectory).}             
\label{tab1}      
\centering                          
\begin{tabular}{c c c c c c c c c c  }        
\hline\hline                 
r & Shock velocity &\multicolumn{2}{c}{Preshock gas} & \multicolumn{2}{c}{Shock Front} & \multicolumn{2}{c}{Adiabatic expansion - start} & \multicolumn{2}{c}{Adiabatic expansion - end} \\    
\hline                        
 (\rstar) & (km s$^{-1})$&T  & n$_{gas}$ & T & n$_{gas}$ & T  & n$_{gas}$& T  & n$_{gas}$  \\
 \hline
   1.2 & 20.0 &2062 & 3.63(13)& 19725 &1.98(14) & 4409& 5.97(14)&1480 & 3.63(13)\\     
   1.5 & 17.9 &1803 & 8.24(12)& 15922 &4.40(13) & 3870& 1.29(14)&1290 & 8.24(12)\\ 
   2 & 15.5 &1517 & 1.44(12)& 12081 &7.59(12) & 3200& 2.14(13)&1080 & 1.44(13)\\ 
   2.5 & 13.9 &1327& 4.24(11)& 9779&2.21(12) &2750& 6.08(12)&951& 4.24(11)\\ 
   3& 12.6& 1190 & 1.69(11)& 8245 &8.73(11) & 2430& 2.35(12)&848 & 1.69(11)\\ 
   4& 11.0 &1001 & 4.51(10)& 6284 &2.29(11) & 1790& 4.48(11)& 711& 4.51(10)\\ 
   5 & 9.8&876 & 1.79(10)&5096 &8.94(10)&1550&1.71(11)&621 &1.79(10)\\
\hline
\end{tabular}   
\tablefoot{ Temperatures T are in Kelvin and gas number densities n$_{gas}$ are in cm$^{-3}$.}                              
\end{table*}

Several formation mechanisms were advocated to explain the presence of H$_2$O in IRC+10216. Melnick et al. (\cite{mel01}) proposed that icy comet bodies orbiting the carbon star could be vaporised in the stellar outflow, providing a source of oxygen to the gas for water formation in the intermediate envelope. Willacy (\cite{wil04}) suggested that water could form on the surface of iron dust grains in the intermediate envelope by Fischer-Tropsch catalysis. Recently, DAB10 and Ag{\'u}ndez et al. (\cite{ag10}, hereafter ACG10) proposed that partial penetration of the interstellar ultraviolet (UV) radiation field occurs deep in the outflow owing to the clumpy nature of the wind. $^{13}$CO and SiO can thus photodissociate, providing atomic oxygen to the gas, which then leads to the formation of H$_2$O. All the proposed explanations for the presence of water have their drawbacks. The first mechanism is somehow extreme because it implies that orbiting icy cometary bodies should be a characteristic of all carbon and S stars. The second proposition requires that iron grains form in the wind of carbon stars, an assumption that still needs confirmation by theoretical models or observations. Finally, according to the third model, all stellar winds are sufficiently clumpy to allow for some penetration of UV photons as deep as 2 \rstar, that is, in the dust formation and wind acceleration region. If so, partial photodissociation of molecular dust precursors, among which the radical propargyl (C$_3$H$_3$) and its precursors CH and CH$_2$, should occur, which would hamper the dust-formation process to a certain degree. 

The formation of the oxygen-bearing species SiO in the inner wind of IRC+12016 was investigated theoretically by Willacy \& Cherchneff (\cite{wil98}, hereafter WC98), who showed that collisional dissociation of CO occurred in the gas layers that experienced the passage of periodic shocks induced by stellar pulsation. The released atomic oxygen then formed a population of OH radicals that triggered the synthesis of SiO via their reaction with atomic Si. Shock-induced chemistry could also explain the formation of CO$_2$ in the O-rich Mira star, IK Tau (Duari et al. \cite{dua99}). In a later study, Cherchneff (\cite{cher06}) modelled the non-equilibrium chemistry of the inner wind of AGB stars as a function of C/O ratios. Of importance was the finding that a few molecules, namely CO, SiO, HCN, and CS, efficiently formed in large amounts in the dust-formation zone whatever the C/O ratio of the star. The author concluded that this group of molecules was ejected as parent species in the intermediate and outer envelopes in O-rich Miras, S stars, and carbon stars. This hypothesis was additionally supported by the observations of high-excitation rotational lines of CO, HCN, SiO, and CS in O-rich, C-rich, and S AGB stars (Decin et al. \cite{dec08}). These results strongly support the assumption that the non-equilibrium chemistry induced by the passage of periodic shocks is responsible for the formation of C-bearing species in the inner envelope of O-rich Miras, and of O-bearing species in the inner wind of carbon stars. 

Here, we revisit the non-equilibirum chemistry of the inner wind of IRC+10216. The updated chemistry includes new processes such as the thermal fragmentation that is active at the high postshock gas temperatures and the formation and destruction of O-bearing and C-bearing species. The latter include acetylene C$_2$H$_2$, hydrocarbons, carbon chains, and the benzene and phenyl aromatic rings, C$_6$H$_6$ and C$_6$H$_5$, respectively. The chemistry of metal hydrides, chlorides, and sulphides, and phosphorous-bearing compounds is also considered. The complete results of this study will be presented in a forthcoming publication, and we report here on our results for water. Section 2 presents the physical and chemical model considered for the inner wind of IRC+10216, while our results for H$_2$O and other important species are summarised in Section 3, and a discussion is presented in Section 4. 

\section{The physical and chemical model}
The dust-formation zone of IRC+12016 extends above the photosphere from 1 to $\sim$ 5 \rstar, according to Keady \& Ridgway (\cite{kea93}). For the sake of clarity, we refer to this region as the 'inner wind' in the rest of the paper. These layers experience the periodic passage of shocks induced by the stellar pulsation. Following Bowen (\cite{bow88}), these shocks compress the gas layers, resulting in high temperatures and densities, which further relax though the collisional dissociation of H$_2$ (Fox \& Wood \cite{fox85}) and adiabatic expansion (Bertschinger \& Chevalier \cite{ber85}). Experiencing the stellar gravitational field, these gas layers fall back to almost their initial position. We model the inner wind following the formalism used by WC98 and Cherchneff (\cite{cher06}), who studied the chemistry of the immediate postshock gas cooling by H$_2$ dissociation and subsequent adiabatic expansion. We assume that the shocks form at a radius r$_s = 1.2$ \rstar\ above the photosphere with an initial strength of 20 km s$^{-1}$ (Ridgway \& Keady \cite{rid81}). The C/O ratio is taken to be equal to 1.4 (Winters et al. \cite{win94}). The initial pre-shock gas density at the shock-formation radius was rescaled down by a factor of ten (n$_{gas}(r_{s})=3.63\times 10^{13}$ cm$^{-3}$) compared to the initial value used by WC98, which was considered to be too high (Ag{\`u}ndez et al. \cite{ag06}). We will see below that this change has no impact on the formed chemical species and their respective abundances. The gas is followed over one pulsation period $P$ from 1 to 5 \rstar, and the preshock and postshock gas parameters as a function of radius are listed in Table~\ref{tab1}. 

\begin{table}
\caption{Chemical species that are linked to the water chemistry and are included in the inner wind model of IRC+10216.}             
\label{tab2}      
\centering                          
\begin{tabular}{l l l l l ll l }        
\hline\hline                 
Atoms&H & O &  C& Si& S & N \\
\hline
Diatomic & H$_2$ & CO & SiO & OH & O$_2$ &SO   \\
&NO &C$_2$ & CS &  CN & CH &SiS\\
&S$_2$& N$_2$ &NH & & & \\
\hline
Tri-atomic & H$_2$O &HCN & CH$_2$ & C$_2$H &HCO &C$_3$ \\
& CO$_2$& NH$_2$& &  & &    \\
\hline
4-atoms&  CH$_3$  & NH$_3$ & C$_2$H$_2$&  & &   \\
\hline
Hydrocarbons &C$_2$H$_3$ & C$_3$H$_2$ & C$_3$H$_3$& C$_4$H$_2$ & C$_4$H$_3$ &C$_4$H$_4$\\
\hline
Aromatics & C$_6$H$_5$& C$_6$H$_6$ & &  &  \\
\hline
\end{tabular}
\end{table}

We updated the chemistry of the inner wind and included all chemical processes relevant  to the high temperatures and number densities characteristic of the shocked gas layers. All chemical pathways leading to the formation of linear molecules, carbon chains, and aromatic rings, include neutral-neutral processes like termolecular, bimolecular reactions and radiative association reactions, whereas destruction is described  by thermal fragmentation and neutral-neutral processes (i.e., oxidation reactions of hydrocarbons and all reverse processes of the formation reactions). No ions are considered in this chemistry because the UV stellar radiation field of IRC+10216 is too low to foster efficient photodissociation and ionisation processes. The periodic shocks also have insufficiently high velocities to be radiative, and as previously mentioned, the immediate postshock gas cools via H$_2$ collisional dissociation (Fox \& Wood \cite{fox85}). The rates for the chemical processes are taken from the National Institute of Standards and Technology database (NIST chemical kinetics database) and also from the literature of combustion, atmopsheric, and material sciences. Species linked to the shock chemistry of H$_2$O in the inner wind  are listed in Table \ref{tab2}, and more detail on the chemistry is given in the online Appendix A.

  \begin{figure}
   \centering
    \includegraphics[width=8.6cm]{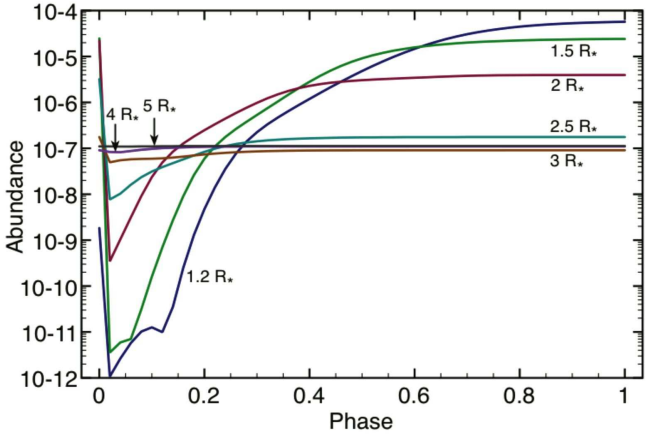}
      \caption{H$_2$O abundances with respect to total gas as a function of pulsation phase $\theta$ and radius in the inner wind (shocks form at $\theta = 0$ and $\theta =1$). }
         \label{fig1}
   \end{figure}

\section{Results}

The abundances of H$_2$O with respect to total gas are presented in Figure \ref{fig1} as a function of radius in the inner wind and pulsation phase $\theta = t / P$ where $t$ is the time and $P$ the pulsation period. The results follow the trends highlighted by our previous studies (WC98, Cherchneff \cite{cher06}). A non-equilibirum chemistry takes place in the post-shock gas at radius 1.2 \rstar, resulting in the collisional breaking of CO at the high post-shock temperatures. H$_2$ is also destroyed in the immediate post-shock gas and reforms in the adiabatic expansion at phase  $\theta >$ 0.3. The released atomic oxygen reacts with H$_2$ to form hydroxyl according to the reaction
 
\begin{equation}
O + H{_2} \longrightarrow OH + H,
\label{eq1}
\end{equation}
 
whose rate has an energy barrier of $\sim$ 3200 K. Therefore, this reaction is efficient at the high post-shock gas temperatures encountered close to the star (see Table \ref{tab1}). Hydroxyl additionally reacts with H$_2$ to form H$_2$O according to
 
\begin{equation}
OH + H{_2} \longrightarrow H{_2}O + H.
\label{eq2}
\end{equation}

\begin{figure}
   \centering
    \includegraphics[width=8.6cm]{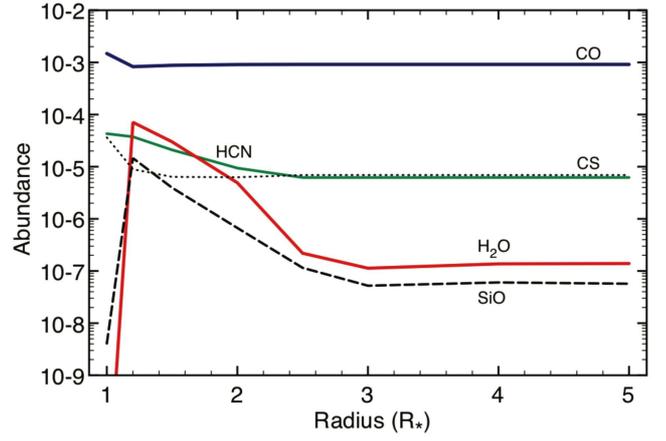}
      \caption{Abundances with respect to H$_2$ of key molecules including H$_2$O taken at $\theta=1$, as a function of radius. Values at $r = 1$ \rstar~are derived from thermodynamical equilibrium calculations.}
         \label{fig2}
   \end{figure}

Reaction \ref{eq2} has a lower energy barrier and a faster rate than Reaction \ref{eq1} and allows for H$_2$O formation at lower temperatures. Close to the star, the hot post-shock gas temperatures preclude the reformation of H$_2$ at early phases and the efficient formation of H$_2$O according to the above processes, leading to very small water abundances. H$_2$O forms at $\theta \geq 0.4$ once H$_2$ has reformed. At radii exceeding 4 \rstar, the shock velocity is low and such that almost no destruction of H$_2$ occurs at early phases. which triggers the early processing of H$_2$O. Once formed, H$_2$O abundances remain constant as the low gas temperatures freeze out the neutral-neutral, high-T chemistry. In Figure \ref{fig2} the abundances with respect to H$_2$ of specific species are shown as a function of radius in the inner wind. These molecules (CO, HCN, SiO, and CS) belong to the group of species predicted to always form with large abundances in the inner wind of stars during their evolution from the Mira to the carbon star stage (Cherchneff \cite{cher06}). The abundance of H$_2$O shows similar trends as that of SiO because those two molecules are competitors in the consumption of OH and their chemistry is coupled (see Appendix A). 

A striking result is the excellent agreement between our predicted water abundances at radii $>$ 3 \rstar~and those derived by Melnick et al. (2001) from their SWAS observations and by DAB10 from their {\it Herschel}/SPIRE/PACS data. At 5 \rstar, we derive an abundance with respect to H$_2$ of $1.4 \times 10^{-7}$ where the values from the SWAS and the {\it Herschel}/HIFI data are $\sim 1 \times 10^{-7}$. In previous studies (WC98, Cherchneff \cite{cher06}), H$_2$O abundances showed similar trends as the present ones, i.e. a very efficient formation at the shock-formation radius and a decrease farther away from the star. However, the decline at larger radii was much sharper than here. This is essentially because of the greater completeness of the chemistry considered in the present model. A more accurate chemistry was considered for silicon and sulphur, leading to lower abundances of SiO than previously found. These SiO abundances with respect to H$_2$ level off at a value of $\sim 6 \times 10^{-8}$ at $r$ = 5 \rstar~and perfectly agree with the value range of $2 \times 10^{-8}-3 \times 10^{-7}$ derived from the new {\it Herschel}/PACs data on Si-bearing species (Decin et al. \cite{dec10}). The lower SiO content results in more available OH radicals to form water, as described in Appendix A. The formation of the aromatic molecule benzene, C$_6$H$_6$, and its radical phenyl, C$_6$H$_5$, as well as their hydrocarbon precursors is specifically taken into account along with their destruction by oxidation reactions. These destruction processes involve OH, the prevalent oxidation agent  in the wind, and result in the formation of water. Finally, in carbon stars, a great fraction of the carbon not locked in CO goes into hydrocarbons, starting with acetylene, C$_2$H$_2$, whose calculated abundances is slightly lower than $1\times 10^{-4}$ in the inner wind. The formation and destruction of C$_2$H$_2$ and larger hydrocarbons impacts on the abundance of molecular hydrogen H$_2$ and indirectly on the formation of water, as explained in Appendix A. These combined effects coupled to the freeze-out of H$_2$O chemistry at r $> 3$ \rstar~lead to higher  H$_2$O abundance values than in previous studies at these radii.

\section{Discussion}
The models of ACG10 and DAB10 allow for the penetration of the interstellar UV radiation field in the very deep layers of the inner wind owing to clumping in the outflow. The models use that of  Ag{\'u}ndez \& Cernicharo (\cite{ag06}), i.e., a Eulerian description of a steady outflow penetrated by some UV radiation (in the form of a minor UV-illuminated wind component), and do not take into account the periodic shocks pervading the inner wind from 1 to $\sim 5$ \rstar. The region exposed to UV radiation is 10 \% of the stellar envelope mass, and the models are highly dependent on the clumping factor and stellar mass-loss. The free oxygen necessary for water synthesis is provided by the partial photodissociation of $^{13}$CO and SiO. This mechanism also triggers the formation of other species of interest, like ammonia. Indeed, NH$_3$ is predicted to form at radius r $\geq$ 3 \rstar~with high abundances peaking at $7\times 10^{-7}$ with respect to H$_2$ at  $r \sim 4 $ \rstar. Keady \& Ridgway (\cite{kea93}) observed mid-IR vibrational transitions of ammonia in IRC+10216 and deduced that a NH$_3$ abundance distribution peaking at 10-20 \rstar~could better reproduce their data. Later, Monnier et al. (\cite{mon00}) carried out interferometric observations of mid-IR molecular absorption bands of NH$_3$ in IRC+10216 with very high spectral resolution. They found that NH$_3$ was located in a region of decaying gas turbulence at radii well beyond the inner wind (r $\geq$ 20 \rstar). These large radii were further used to model NH$_3$ submm lines observed with the satellite ODIN (Hagesawa et al. \cite{hag06}). The above-mentioned observations suggest that the formation locus of NH$_3$ is located well beyond the dust-formation zone (r $\geq$ 5 \rstar), while ammonia is formed as early as 3 \rstar with a high abundance extending to $\sim$100 \rstar ~according to DAB10 and ACG10. Our present model does not form NH$_3$ in the inner wind ($x$(NH$_3$)  $\sim 4\times 10^{-13}$ with respect to H$_2$ at r = 5 \rstar), supporting larger formation radii for this species in accordance with mid-IR and submm observations. 

Decin et al. ({\cite{dec110}, DAB10) report on newly observed high-excitation lines of cyanoacetylene HC$_3$N with the IRAM telescope. These high-J lines present a flat-topped profile, and DAB10 and ACG10 construe this shape as evidence for a formation locus deeper in the envelope. They claim that the dissociation occurring in their minor UV-illuminated wind component explains the formation of HC$_3$N in the intermediate envelope. An enhancement (up to $\sim 3 \times 10^{-7}$) in the HC$_3$N abundance peaking at $r= 120$ \rstar~is required to reproduce the high-J IRAM lines. The radial distribution of HC$_3$N in IRC+10216 was mapped by Audinos et al. (\cite{au94}) with the IRAM telescope, and the line modelling for a steady homogeneous outflow already required an enhanced HC$_3$N abundance at these radii. Cherchneff \& Glassgold (\cite{cher93}) modelled the cyanopolyyne chemistry in IRC+10216 and found a shoulder in the HC$_3$N distribution at radii $<$ 200 \rstar~resulting from the synthesis of HC$_3$N by neutral-neutral channels. The predicted HC$_3$N abundances in the shoulder were a factor $\sim 7$ lower compared to values derived by Audinos et al. We notice that the HC$_3$N abundance distribution of DAB10 for their major UV-shielded component does not include this predicted shoulder, which points to a different chemistry used by DAB10 and ACG10. High angular resolution observations by Trung \& Lim (\cite{tru08}) of the J = 5 - 4 HC$_3$N line with the VLA  clearly indicate a very clumpy shell distribution as far as 20'' from the star. Therefore, radiative transfer models taking into account the outflow clumpiness are required to accurately assess the HC$_3$N abundance distribution across the envelope. An updated chemical model is also necessary to clearly identify the prevalent formation pathways to HC$_3$N, including neutral-neutral and UV dissociation-induced processes. A shock-induced chemistry may also be a viable source of HC$_3$N at intermediate radii and must be considered in subsequent models. 

The present results support the hypothesis that H$_2$O forms very close to the star with high abundances with respect to H$_2$ between $1 \times 10^{-6}$ and $1 \times 10^{-4}$ and that the abundance gradually chemically freezes out to a value of $1.4 \times 10^{-7}$ at r $\geq$ 5 \rstar. The contribution to the H$_2$O line intensities from the high-abundance region comprised between 1.2 \rstar~and 2.5 \rstar~is not observable with {\it Herschel} because of its small extent and beam filling factor. Furthermore, the H$_2$O formation chemistry appears to be coupled to that of SiO and highlights the importance of the hydroxyl radical OH in the inner wind. A major advantage of the shock-induced chemistry hypothesis is that any star on the AGB pulsates and experiences the passage of shocks in its dust-formation zone. Therefore, it provides a non-restrictive, genuine mechanism for forming water and other molecules very close to the star for a variety of objects in different evolutionary stages, potentially explaining the detection of H$_2$O in O-rich, S and carbon stars with the {\it Herschel} telescope. The shock-induced chemistry is also successful in explaining the presence of O-bearing molecules in carbon stars and C-bearing species in O-rich Miras. Foremost, this hypothesis does not exclude other formation mechanisms for water at larger radii (comets, chemistry on grain surfaces, partial photodissociation in a clumpy outflow)\footnotemark. Because of its presence in AGB stars, H$_2$O can be added to the list of species (CO, SiO, HCN, CS) proposed by Cherchneff (\cite{cher06}) to efficiently form in the dust-formation zone from shock chemistry and be ejected as 'parent' species in the intermediate and outer envelopes. This hypothesis awaits testing by observations of the deep layers of AGB envelopes at mid-IR and submm wavelengths. 

\footnotetext{New data on water in IRC+10216 from {\it Herschel}/HIFI have just been released (Neufeld et al. \cite{neu101}) and confirm the formation of H$_2$O at radii $\leq$ 6 \rstar. They indicate a common locus for H$_2$O and SiO formation in agreement with the present results and rule out the comet and Fischer-Tropsch catalysis scenarios for H$_2$O formation. This is further supported by new {\it Herschel}/HIFI data for several carbon stars, confirming the widespread presence of water very close to the star (Neufeld et al. \cite{neu102}).} 

\begin{acknowledgements}
The author thanks the anonymous referee  for helpful comments on how to improve the manuscript and Dinh-V-Trung and Alexander Tielens for stimulating discussions. 
\end{acknowledgements}

\Online 
\begin{appendix} 
\section{Chemical model and new key processes} 

\begin{table}
\caption{Chemical reaction types included in the chemical model of IRC+10216 inner wind.}             
\label{tab3}      
\centering                          
\begin{tabular}{l l l}        
\hline\hline                 
Unimolecular & AB $\longrightarrow$ A + B & Decomposition \\
\hline
Bimolecular& A + B $\longrightarrow$ C + D & Neutral-neutral \\
 &A + B $\longrightarrow$ AB + $h\nu$ & Radiative association  \\
  & AB + M $\longrightarrow$ A + B + M & Thermal fragmentation  \\
\hline
Termolecular & A + B + M $\longrightarrow$ AB + M & Three-body association \\
\hline
\end{tabular}
\end{table}

The present chemical model is based on that of Willacy \& Cherchneff ({\cite{wil98}}, hereafter WC98) and includes all processes relevant to the hot and dense gas pertaining to the inner wind between 1 and 5 \rstar. However, many processes and their rates have been updated and/or added in view of the progresses made in chemical kinetics, and combustion and aerosol chemistry since 1998. The various types of chemical pathways considered in the model are summarised in Table~\ref{tab3}. Thermal fragmentation, i.e., destruction of molecules by collision with the ambient gas, unimolecular decomposition of aromatics and hydrocarbons, and radiative association reactions were added to the chemistry. The chemical network includes 59 species (some of which are listed in Table \ref{tab2}) and 370 chemical reactions. The rates for these chemical processes are taken from the National Institute of Standards and Technology database (NIST chemical kinetics database), and from the literature of combustion, atmopsheric, and material sciences. 

In terms of formalism, three major changes were implemented with respect to WC98 and Cherchneff (\cite{cher06}). Firstly, the treatment of the reverse reaction of a specific process was changed. Several new rates were measured in combustion and aerosol chemistry and are now available. Therefore, instead of calculating the rate of the reverse process from detailed balance and the equilibrium constant (see Equation 4 in WC98), we directly enter the available measured or calculated rate values in the chemical network. When the information is not available, we make 'educated' guesses depending on the type of the reaction. 

Secondly, the chemistry involving atomic silicon, Si, and Si-bearing species has been restricted to reactions for which rates were measured or calculated. The reactions and rates derived from the isovalence of Si with carbon as stated in WC98 were abandoned. The prevalent formation reactions for SiO are
\begin{equation}
Si + CO \longrightarrow SiO + C,
\label{A1}
\end{equation}
and
\begin{equation}
Si + OH \longrightarrow SiO + H,
\label{A2}
\end{equation}

Reaction \ref{A1} is effective at forming SiO at temperatures $\geq$ 3000 K, whereas Reaction \ref{A2} contributes to most of the SiO synthesis at temperatures $<$ 3000 K. Destruction of SiO is mainly triggered by the opposite process of Reaction \ref{A2} at all temperatures. Reaction \ref{A2} is a competitive channel to the formation of H$_2$O  via Reaction \ref{eq2}, because it depletes OH radicals. But because of the high H$_2$ content of the wind, the net formation rate of H$_2$O according to Reaction \ref{eq2} is much higher than that for SiO. Furthermore, a large quantity of atomic Si is tied up in silicon sulphide, SiS, leaving a reduced pool of available Si to form SiO. These combined effects result in water abundances higher than those of SiO, as seen in Figure \ref{fig2}. 

Thirdly, the chemistry now included the formation of a larger set of molecules, some of them listed in Table~\ref{tab2}. As explained in Section 3, the formation of single aromatic ring compounds (benzene and phenyl) was considered. Previous studies of the formation of polycyclic aromatic hydrocarbons in AGB winds never took into account the presence of O-bearing species in the shocked regions (Cherchneff \cite{cher10}). The coupling of these C- and O-rich chemistries is important, because the unimolecular decomposition of hydrocarbons and aromatics replenish the gas in acetylene. Since C$_2$H$_2$ has large abundances in the inner wind and its formation and destruction processes are coupled to the H$_2$ chemistry, it indirectly impacts on the formation of OH, H$_2$O, and SiO through Reactions \ref{eq1}, \ref{eq2}, and \ref{A2}. Furthermore, the potential oxidation of hydrocarbon precursors by O-bearing species, namely OH, is included and gives back H$_2$O to the gas phase, albeit its impact is minor in replenishing water.

\end{appendix}

\end{document}